\documentclass[prd,showpacs]{revtex4}
\usepackage[dvips]{epsfig}
\usepackage{color}
\def\beq{\begin{equation}}
\def\eeq{\end{equation}}
\def\bea{\begin{eqnarray}}
\def\eea{\end{eqnarray}}
\def\bq{\begin{quote}}
\def\eq{\end{quote}}

\def\la{\langle}
\def\ra{\rangle}
\def\nin{\noindent}
\def\ba{\vspace*{-0.2cm}\begin{array}}
\def\ea{\end{array}\vspace*{-0.2cm}}

\def\als{\alpha_s}

\def\gg2{ \la\alpha_s G2 \ra}
\def\gg3{g^3f_{abc}\la G^aG^bG^c \ra}
\def\ggg4{\la\als^2G4\ra}

\def\beq{\begin{equation}}
\def\enq{\end{equation}}
\def\beqa{\begin{eqnarray}}
\def\enqa{\end{eqnarray}}

\def\MeV{\nobreak\,\mbox{MeV}}
\def\GeV{\nobreak\,\mbox{GeV}}

\def\qq{\lag\bar{q}q\rag}

\def\ss{\lag\bar{s}s\rag}
\def\mix{\lag\bar{q}g\si.Gq\rag}
\def\mixs{\lag\bar{s}g\si.Gs\rag}

\def\G3{\lag g^3G3\rag}

\def\pli{p^\prime}

\def\si{\sigma}

\def\al{\alpha}

\def\lb{\label}
\def\nn{\nonumber}

\newcommand{\rag}{\rangle}
\newcommand{\lag}{\langle}

\begin{document}

\title{Predicition for the decay width of a charged state near the
$D_s\bar{D}^*/D^*_s\bar{D}$ threshold}

\author{Jorgivan M. Dias$^1$}
\email{jdias@if.usp.br}
\author{Xiang Liu$^{2,3}$}
\email{xiangliu@lzu.edu.cn}
\author{Marina Nielsen$^1$}
\email{mnielsen@if.usp.br}
\affiliation{$^1$Instituto de F\'{\i}sica, Universidade de S\~{a}o Paulo,
C.P. 66318, 05389-970 S\~{a}o Paulo, SP, Brazil\\
$^2$Research Center for Hadron and CSR Physics,
Lanzhou University and Institute of Modern Physics of CAS, Lanzhou 730000, China\\
$^3$School of Physical Science and Technology, Lanzhou University,
Lanzhou 730000,  China}

\begin{abstract}
\nin
Very recently it was predicted the existence of a charged state near the
$D_s\bar{D}^*/D^*_s\bar{D}$ threshold. This state, that we call $Z_{cs}^+$,
would be the strange partner of the recently observed $Z_c^\pm(3900)$. Using
standard techniques of QCD sum rules, we evaluate the three-point function for
the vertices $Z_{cs}^+ \, J/\psi \, K^+$, $Z_{cs}^+ \, \eta_c \, K^{*+}$ and
$Z_{cs}^+ \, D_s^+\bar{D}^{*0}$ and we make predictions for the corresponding
decay widths in these channels.
\end{abstract}
\pacs{ 11.55.Hx, 12.38.Lg , 12.39.-x}
\maketitle


In a pioneering work, using the initial single pion emission mechanism (ISPE),
the authors of ref.~\cite{Chen:2011xk} have predicted the existence of a
charged state, close to the $D^*\bar{D}$ threshold, in the hidden-charm
dipion decay of the charmonium-like structure $Y(4260)$. This state,
called $Z_c^+(3900)$, was soon after observed by the BESIII and BELLE
collaborations  in $e^+e^-\to J/\psi\pi^+\pi^-$ at $\sqrt{s}=4260~\MeV$
\cite{Ablikim:2013mio,Liu:2013dau}. This observation was also confirmed by the
authors of ref.~\cite{Xiao:2013iha} using  CLEO-c data.
Stimulated by this discovery, the authors of ref.~\cite{Chen:2013wca}
have extended  the ISPE mechanism to include the kaon, the chiral partner of
the pion. They call it the initial single chiral particle emission (ISChE)
mechanism. Under the ISChE mechanism it is possible to study the hidden-charm
dikaon decay of  a charmonium-like states. In particular, studying the
hidden-charm dikaon decay of  the charmonium-like structure $Y(4660)$, the
authors of ref.~\cite{Chen:2013wca} find a sharp peak structure close to the
$D_s\bar{D}^*/D^*_s\bar{D}$ threshold. Therefore, a charged charmonium-like
structure with hidden-charm and open-strange channels with mass close to the
$D_s\bar{D}^*/D^*_s\bar{D}$ threshold, which we call $Z_{cs}^\pm$,
should be seen in the $Y(4600)\to J/\psi K^+K^-$ decay.


The mass of a $J^P=1^+$ $D_s\bar{D}^*$ molecular state was first predicted,
using the QCD  sum rules (QCDSR) method \cite{svz,rry,SNB}, in
ref.~\cite{Lee:2008uy}. They found $m_{Z_{cs}}=(3.97\pm0.08)~\GeV$, which is
very close to the $D_s^+\bar{D}^{*0}$ threshold at 3.976 GeV.
In this work we use the  method of QCDSR to study some hadronic decays
of $Z_{cs}^\pm$, considering the $Z_{cs}$ as a tetraquark state, similar to what
was done for the $Z_c^\pm(3900)$ state in ref.~\cite{Dias:2013xfa}.
Therefore, the interpolating field for $Z_{cs}^+$ is given by:
\beq
j_\alpha={i\epsilon_{abc}\epsilon_{dec}\over\sqrt{2}}[(u_a^TC\gamma_5c_b)
(\bar{s}_d\gamma_\alpha C\bar{c}_e^T)-(u_a^TC\gamma_\alpha c_b)
(\bar{s}_d\gamma_5C\bar{c}_e^T)]\;,
\label{field}
\enq
where $a,~b,~c,~...$ are color indices, and $C$ is the charge conjugation
matrix. The mass obtained in QCDSR for the
$Z_{cs}$ state described by the current in Eq.~(\ref{field}) is the same
as the one obtained in  \cite{Lee:2008uy}, as expected from the
results presented in ref.~\cite{Narison:2010pd}. Therefore, here we evaluate
only the decay width. For a comprehensive review of
the use of different currents to describe four-quark states we refer the reader
to \cite{review}.

We will consider four decay channels: $Z_{cs}^+ \to  J/\psi \, K^+$,
$Z_{cs}^+ \to  \eta_c \, K^{*+}$,  $Z_{cs}^+ \to   \bar{D}^{*0}\, D_s^+$
and $Z_{cs}^+ \to   \bar{D}^{0}\, D_s^{*+}$. Besides these four discussed 
decay channels, $Z_{cs}^+\to \chi_{c0}K^+$ via P-wave is allowed, where the sum
  of the masses of $\chi_{c0}$ and Kaon is about 3912 MeV less than the central
 value of the mass of $Z_{cs}^+$ \cite{Lee:2008uy}. However, in this work we 
will not include this channel in our discussion since this P-wave decay and 
small phase space can suppress the decay width of $Z_{cs}^+\to \chi_{c0}K^+$ 
compared with these two S-wave hidden-charm decay channels $Z_{cs}^+ \to  
J/\psi \, K^+$ and $Z_{cs}^+ \to  \eta_c \, K^{*+}$.

In these four channels there is always a vector and a pseudoscalar mesons
as final states. For the last three cases the pseudoscalar mesons are described by
pseudoscalars currents:
\beq
j_{5}^{\eta_c}=i\bar{c}_a\gamma_5 c_a,\;j_{5}^{D}=i\bar{c}_a\gamma_5 u_a,\;
\mbox{ and }\; j_{5}^{D_s}=i\bar{s}_a\gamma_5 c_a.
\lb{pseudo}
\enq
However, it is well known that the kaon can not be well described, in QCDSR,
by a pseudoscalar current \cite{Novikov:1981xi}. Therefore, in the case of the
$Z_{cs}^+ \to  J/\psi \, K^+$ decay, we use an axial current to describe the kaon
\beq
j_{5\nu}^{K}=\bar{s}_a\gamma_5\gamma_\nu u_a.
\lb{kaon}
\enq
For the vector mesons we use the currents
\beq
j_{\mu}^{\psi}=\bar{c}_a\gamma_\mu c_a,\;\;j_{\mu}^{D^*}=\bar{c}_a\gamma_\mu
u_a,\;\;
j_{\mu}^{D_s^*}=\bar{s}_a\gamma_\mu c_a\;\mbox{ and }\;
j_{\mu}^{K^*}=\bar{s}_a\gamma_\mu u_a.
\lb{vector}
\enq

The QCDSR calculation of these four vertices are based on
the three-point function given by:
\beq
\Pi_{\mu i\al}(p,\pli,q)=\int d^4x~ d^4y ~e^{i\pli.x}~e^{iq.y}~
\Pi_{\mu i\al}(x,y),
\lb{3po}
\enq
with
\beqa
\Pi_{\mu\nu\al}(x,y)&=&\lag 0 |T[j_\mu^{\psi}(x)j_{5\nu}^{K}(y)
j_\alpha^\dagger(0)]|0\rag,\nn\\
\Pi_{\mu\al}(x,y)&=&\lag 0 |T[j_5^{\eta_c}(x)j_{\mu}^{K^*}(y)
j_\alpha^\dagger(0)]|0\rag,\nn\\
\Pi_{\mu\al}(x,y)&=&\lag 0 |T[j_\mu^{D^*}(x)j_{5}^{D_s}(y)
j_\alpha^\dagger(0)]|0\rag,\nn\\
\Pi_{\mu\al}(x,y)&=&\lag 0 |T[j_\mu^{D_s^*}(x)j_{5}^{D}(y)
j_\alpha^\dagger(0)]|0\rag,
\label{corr}
\enqa
for the four decays. In Eq.(\ref{corr}) $p=\pli+q$.

\begin{figure}[h]
\centerline{\epsfig{figure=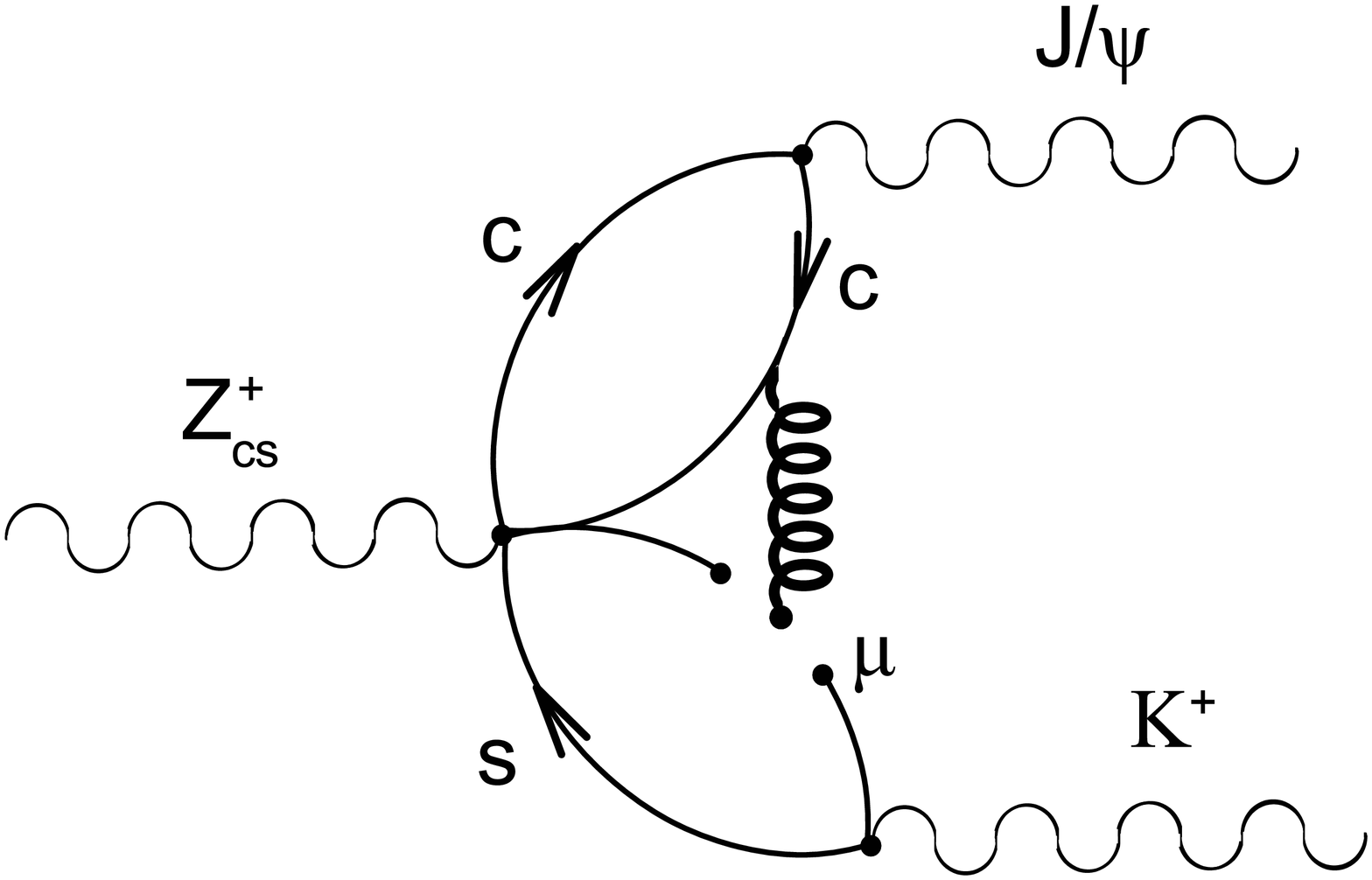,height=50mm}}
\caption{CC diagram which contributes to the OPE side of the sum rule.}
 \label{fig1}
\end{figure}

To assure that the non-trivial color structure of the
current in Eq.~(\ref{field}) is maintained in the QCDSR calculation, in the OPE
side we will consider only the diagrams with non-trivial color structure,
as in ref.~\cite{Dias:2013xfa}. These diagrams are called color-connected (CC)
diagrams. In the case of the $Z_{cs}^+\to J/\psi K^+$ decay, one of the CC
diagrams that contribute to the OPE side is shown  in Fig. \ref{fig1}.  Possible
permutations (not shown) of the diagram in Fig.~\ref{fig1} also contribute.

The diagram in Fig.~\ref{fig1} contributes to many structures. However,
as we can see below, only the structures $q_\nu g_{\mu\alpha}$ and
$q_\nu\pli_\mu\pli_\alpha$ also appear in the phenomenological side. Following
\cite{Dias:2013xfa} we choose to work with the $q_\nu\pli_\mu\pli_\alpha$
structure. Therefore  in the OPE side and in
the $q_\nu\pli_\mu\pli_\alpha$ structure we obtain:
\beq
\Pi^{(OPE)}= {(\mix+\mixs)\over24\sqrt{2}\pi^2}{1\over q^2}
\int_0^1 d\alpha{\alpha(1-\al)\over m_c^2-\al(1-\al){\pli}^2}.
\label{ope}
\enq

The phenomenological side of the sum rule can be evaluated by inserting
intermediate states for $Z_{cs}$, $J/\psi$ and $K$ into Eq.(\ref{3po}).
We get:
\beq
\Pi_{\mu\nu\al}^{(phen)} (p,\pli,q)={\lambda_{Z_{cs}} m_{\psi}f_{\psi}F_{K}~
g_{Z_{cs}\psi K}(q^2)q_\nu
\over(p^2-m_{Z_{cs}}^2)({\pli}^2-m_{\psi}^2)(q^2-m_K^2)}
~\left(-g_{\mu\lambda}+{\pli_\mu \pli_\lambda\over m_{\psi}^2}\right)
\left(-g_\alpha^\lambda+{p_\alpha p^\lambda\over m_{Z_{cs}}^2}\right)
+\cdots\;.
\lb{phen}
\enq
The contribution of the excited states are included by the dots. These include
pole-continuum and continuum contributions. The form factor, $g_{Z_{cs}\psi K}
(q^2)$, appearing in Eq.~(\ref{phen}), is defined as the generalization for a
off-shell kaon, of the on-mass-shell coupling constant $g_{Z_{cs}\psi K}$. The
coupling constant can be extracted from the effective lagrangian
\beq
{\cal{L}}=g_{Z_{cs}\psi K}{Z}_{cs}^\mu\psi_\mu\bar{K} + cc.
\label{lag}
\enq
From the lagrangian in Eq.~(\ref{lag}) we get:
\beq
\lag J/\psi(\pli) K(q)|Z_{cs}(p)\rag=g_{Z_{cs}\psi K}(q^2)
\varepsilon^*_\lambda(\pli)\varepsilon^\lambda(p),
\label{coup}
\enq
where
$\varepsilon_\alpha(p),~\varepsilon_\mu(\pli)$ are  the polarization
vectors of the $Z_{cs}$ and $J/\psi$ mesons  respectively.

The coupling $\lambda_{Z_{cs}}$ and the meson decay constants
$f_\psi$ and $F_K$ appearing in Eq.~(\ref{phen}) are defined through the
current-state couplings:
\beqa
\lag 0 | j_\mu^\psi|J/\psi(\pli)\rag &=&m_\psi f_{\psi}\varepsilon_\mu(\pli),
\nn\\
\lag 0 | j_{5\nu}^K|K(q)\rag &=&iq_\nu F_K,
\nn\\
\lag Z_{cs}(p) | j_\alpha|0\rag &=&\lambda_{Z_{cs}}\varepsilon_\al^*(p).
\lb{fp}
\enqa

If one neglects the kaon mass in the right hand side of Eq.~(\ref{phen}) we can
extract directly the coupling constant, $g_{Z_{cs}\psi \pi}$, instead
of the form factor, like in \cite{Bracco:2011pg,Dias:2013xfa}. Therefore,
isolating the $q_\nu\pli_\mu\pli_\alpha$ structure in Eq.~(\ref{phen}) and
making a single Borel transformation to both $P^2={P^\prime}^2\rightarrow M^2$,
we get the sum rule:
\beq
A\left(e^{-m_\psi^2/M^2}-e^{-m_{Z_{cs}}^2/M^2}\right)+B~e^{-s_0/M^2}
={(\mix+\mixs)\over24\sqrt{2}\pi^2}
\int_0^1 d\alpha \,  e^{- m_c^2\over \al(1-\al)M^2},
\label{sr}
\enq
where $s_0$ is the continuum threshold parameter for $Z_{cs}$: $\sqrt{s_0}=
(4.5\pm0.1)~\GeV$ \cite{Lee:2008uy}, and
\beq
A={g_{Z_{cs}\psi K}\lambda_{Z_{cs}} f_{\psi}F_{K}~(m_{Z_{cs}}^2+m_\psi^2+m_K^2)
\over 2m_{Z_{cs}}^2m_{\psi}(m_{Z_{cs}}^2-m_{\psi}^2)}.
\label{a}
\enq

As commented above, the dots in Eq.~(\ref{phen}) include
pole-continuum and continuum contributions.
The parameter $B$ in Eq.(\ref{sr}) is introduced to take into account the
contributions associated with pole-continuum transitions, which
are not suppressed  when only a single Borel transformation is done in a
three-point function sum rule, as shown in \cite{Navarra:2006nd,col,bel,io1}.

The numerical values for quark masses and QCD condensates used in this
calculation are  listed in Table \ref{TabParam} \cite{SNB,narpdg}.

{\small
\begin{table}[h]
\setlength{\tabcolsep}{1.25pc}
\caption{QCD input parameters.}
\begin{tabular}{ll}
&\\
\hline
Parameters&Values\\
\hline
$m_c$ & $(1.18 - 1.28) \GeV$ \\
$\qq$ & $-(0.23 \pm 0.03)^3\GeV^3$\\
$m_0^2 \equiv \mix/ \qq$ & $(0.8 \pm 0.1) \GeV^2$\\
$\ss/ \qq$ & $0.8$ \\
\hline
\end{tabular}
\label{TabParam}
\end{table}}

The numerical values of the meson masses and decay constants used in all
calculations  are given in Table \ref{TabHadron}.

{\small
\begin{table}[h]
\setlength{\tabcolsep}{1.25pc}
\caption{Meson masses and decay constants.}
\begin{tabular}{lll}
&\\
\hline
Quantity&Value&Ref.\\
\hline
$m_\psi$ & $3.1 \GeV$ & \cite{pdg}\\
$m_{\eta_c}$ & $2.98~\GeV$ & \cite{pdg}\\
$m_{D^*}$ & $2.01~\GeV$ & \cite{pdg}\\
$m_{D_s^*}$ & $2.11~\GeV$ & \cite{pdg}\\
$m_{D_s}$ & $1.97~\GeV$ & \cite{pdg}\\
$m_{D}$ & $1.87~\GeV$ & \cite{pdg}\\
$m_K^*$ & $0.892\GeV$ & \cite{pdg}\\
$m_K$ & $0.494\GeV$ & \cite{pdg}\\
$f_{\psi}$ & $0.405 ~\GeV$ & \cite{pdg}\\
$f_{\eta_c}$ & $0.35~\GeV$ & \cite{nov}\\
$f_{D_s^*}$ & $0.33~\GeV$ & \cite{borges}\\
$f_{D_s}$ & $(0.24\pm0.08)~\GeV$ & \cite{Blossier:2009bx}\\
$f_{D^*}$ & $(0.24\pm0.02)~\GeV$ & \cite{Bracco:2011pg}\\
$f_{D}$ & $(0.18\pm0.02)~\GeV$ & \cite{Bracco:2011pg}\\
$f_K$ & $(0.16\pm0.02)~\GeV$ &\cite{pdg}\\
$f_K^*$ & $(0.22\pm0.01)~\GeV$ &\cite{pdg}\\
\end{tabular}
\label{TabHadron}
\end{table}}
For the $Z_{cs}$ mass and the meson-current coupling,
$\lambda_{Z_{cs}}$, defined in Eq.(\ref{fp}), we use the values determined from
the two-point sum rule \cite{Lee:2008uy}: $m_{Z_{cs}}=(3.97\pm0.08)~\GeV$ and
$\lambda_{Z_{cs}}=(1.8\pm0.2)\times10^{-2}~\GeV^5$.

\begin{figure}[h]
\centerline{\epsfig{figure=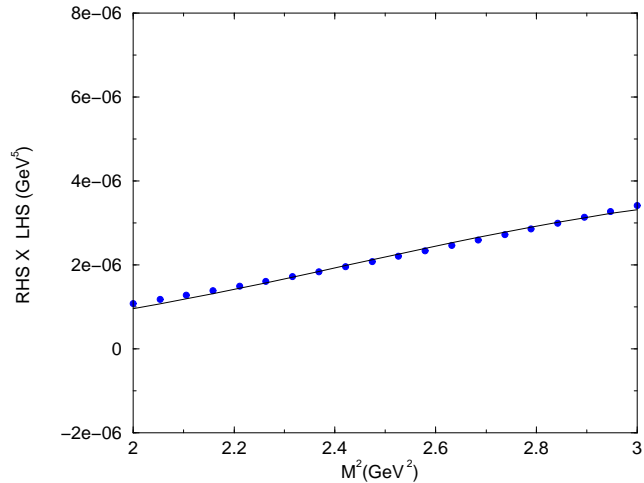,height=65mm}}
\caption{Dots: the RHS of Eq.(\ref{sr}), as a function of the Borel mass
for $\sqrt{s_0}=4.5$ GeV.
The solid line gives the fit of the QCDSR results through
the LHS of Eq.(\ref{sr}).}
\label{fig2}
\end{figure}

In \cite{Lee:2008uy} it was shown that the Borel window where the two-point
function for $Z_{cs}$  shows good OPE convergence and  pole dominance
is in the range$2.0 \leq M^2 \leq 3.0$ GeV$^2$. Therefore, we use here this same
Borel window.
In Fig.~\ref{fig2} we show, through the circles, the right-hand side (RHS) of
Eq.(\ref{sr}), {\it i.e.} the OPE side of the sum rule, as a function of the
Borel mass. We can fit the OPE results with the analytical expression in the
left-hand side (LHS) of Eq.(\ref{sr}). We get: $A=1.28\times10^{-4}~\GeV^5$ and
$B=-1.03\times10^{-3}~\GeV^5$, using $\sqrt{s_0}=4.5~\GeV$. Using the value
obtained for $A$ through the fit and the expression in Eq.(\ref{a}) we get for
coupling constant: $g_{Z_{cs}\psi K}=2.57~\GeV$.
Considering the uncertainties given in the parameters in Tables I and II,
we obtain:
\beq
g_{Z_{cs}\psi K}=(2.58\pm0.30)~\GeV.
\label{coupling}
\enq
With the value of $g_{Z_{cs}\psi K}$ we can estimate the decay width using the
expression \cite{Dias:2013xfa}:
\beq
\Gamma(Z_{cs}^+\to J/\psi K^+)={p^*(m_{Z_{cs}},m_\psi,m_K)\over8\pi m_{Z_{cs}}^2}
{1\over3}g^2_{Z_{cs}\psi K}\left(
3+{(p^*(m_{Z_{cs}},m_\psi,m_K))^2\over m_{\psi}^2}\right),
\label{dec}
\enq
where
\beq
p^*(a,b,c)={\sqrt{a^4+b^4+c^4-2a^2b^2-2a^2c^2-2b^2c^2}\over 2a}.
\enq
Here, the mass of $Z_{cs}^+$ is taken as $(3.97\pm 0.08)$ GeV, which is from the 
QSR calculation \cite{Lee:2008uy}. We obtain:
\beq
\Gamma(Z_{cs}^+\to J/\psi K^+)=(11.2\pm3.5)~\MeV.
\label{width}
\enq

One can notice that the coupling in this case is smaller than $g_{Z_{c}\psi \pi}$,
obtained in \cite{Dias:2013xfa}. One of the possible reasons for that is the fact
that the OPE side, in the $Z_{cs}$ case, is smaller than the corresponding one
for $Z_c^+(3900)$, due to the presence of the strange-quark condensate. Also,
the current-coupling parameter $\lambda_{Z_{cs}}$ is bigger than $\lambda_{Z_c}$.
In addition, the phase space of $Z_{cs}^+$ decay into $J/\psi K^+$ is smaller than
that of  $Z_c^+(3900)\to J/\psi \pi)$, which is a reason why
$\Gamma(Z_{cs}^+\to J/\psi K^+)$ is less than half of the
$\Gamma(Z_c^+(3900)\to J/\psi \pi)$.


Let us consider now the $Z_{cs}^+\to \eta_c \, K^{*+}$ decay.
Considering only CC diagrams, like the one in Fig.~\ref{fig1}, we get for
the OPE side in the $\pli_\mu q_\alpha$ structure:
\beq
\Pi^{(OPE)}= {-i m_c(\mix+\mixs)\over96\sqrt{2}\pi^2}{1\over q^2}
\int_0^1 d\alpha{1\over m_c^2-\al(1-\al){\pli}^2}.
\label{ope2}
\enq

The phenomenological side is obtained by saturating the correlation function in
Eq.~(\ref{3po}) with $Z_{cs}^+,~ \eta_c$ and $K^{*+}$ states. The decay
constants for vector ($V$) and pseudocalar ($P$) states are defined through
the coupling of the current with the states:
\beqa
\lag 0 | j_\mu^V|V(q)\rag &=&m_V f_{V}\varepsilon_\mu(q),
\nn\\
\lag 0 | j_{5}^{P}|P(q)\rag &=& {f_{P}m^2_{P}\over m_{q_1}+m_{q_2}},
\lb{fp2}
\enqa
where $m_{q_1}$ and $m_{q_2}$ are the masses of the constituents quarks of the
pseudoscalar meson $P$.

We get for the phenomenological side
\beq
\Pi_{\mu\al}^{(phen)} (p,\pli,q)={-i\lambda_{Z_{cs}} m_{K^*}f_{K^*}f_{\eta_c}
m^2_{\eta_c}~g_{Z_{cs}\eta_c K^*}(q^2)
\over2m_c(p^2-m_{Z_{cs}}^2)({\pli}^2-m_{\eta_c}^2)(q^2-m_{K^*}^2)}
\left(-g_{\mu\lambda}+{q_\mu q_\lambda\over m_{\rho}^2}\right)
\left(-g_\alpha^\lambda+{p_\alpha p^\lambda\over m_{Z_c}^2}\right)
+\cdots.
\lb{phen2}
\enq
Isolating the $q_\al\pli_\mu$ structure in Eq.~(\ref{phen2}) and
making a single Borel transformation on  both $P^2={P^\prime}^2$,
we get:
\beq
C\left(e^{-m_{\eta_c}^2/M^2}-e^{-m_{Z_{cs}}^2/M^2}\right)+D~e^{-s_0/M^2}=
{Q^2+m_\rho^2\over Q^2}{m_c(\mix+\mixs)\over96\sqrt{2}\pi^2}
\int_0^1 d\alpha {e^{- m_c^2\over \al(1-\al)M^2}\over\al(1-\al)},
\label{sr2}
\enq
where $Q^2=-q^2$ and the parameter $C$ is given in terms of the form factor:
\beq
C={g_{Z_{cs}\eta_c K^*}(Q^2)\lambda_{Z_{cs}} m_{K^*} f_{K^*}f_{\eta_c}m_{\eta_c}^2
\over2m_c m_{Z_{cs}}^2(m_{Z_{cs}}^2-m_{\eta_c}^2)}.
\label{c}
\enq

\begin{figure}[h]
\centerline{\epsfig{figure=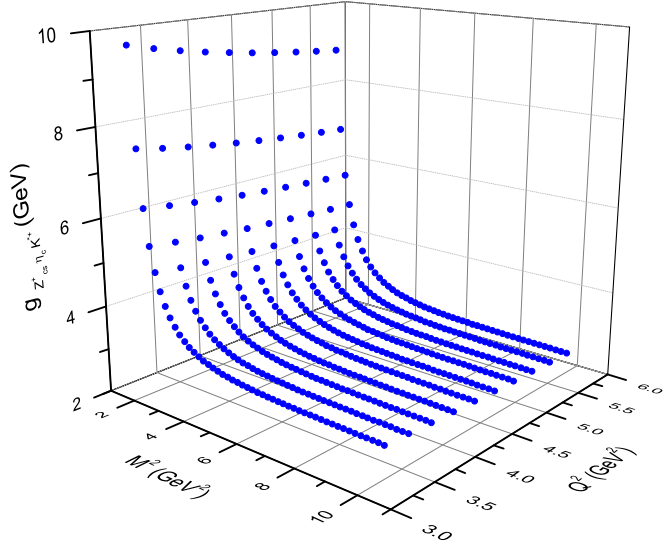,height=80mm}}
\caption{QCDSR results for the form factor  $g_{Z_{cs}\eta_c K^*}(Q^2)$ as
a function of $Q^2$ and $M^2$ for $\sqrt{s_0}=4.5$ GeV.}
\label{fig3}
\end{figure}
To determine $ g_{Z_{cs}\eta_c K^*}(Q^2)$ we use Eq.~(\ref{sr2}) and its
derivative with respect to $M^2$ to eliminate $D$ from Eq.~(\ref{sr2}).
The form factor $g_{Z_{cs}\eta_c K^*}(Q^2)$ is shown in Fig.~\ref{fig3}, as a
function of both $M^2$ and $Q^2$. To extract $g_{Z_{cs}\eta_c K^*}(Q^2)$ we need
first to establish the Borel window where the sum rule is as much independent of
the Borel mass as possible. From Fig. \ref{fig3} we notice that this happens
in the region $4.0 \leq M^2 \leq 10.0$ GeV$^2$.

\begin{figure}[h]
\centerline{\epsfig{figure=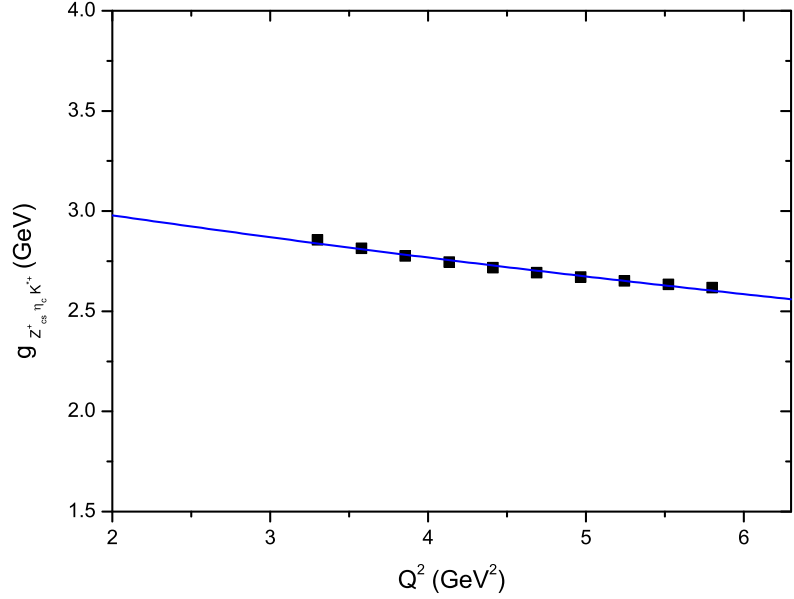,height=80mm}}
\caption{QCDSR results for $g_{Z_{cs}\eta_c K^*}(Q^2)$, as a function of $Q^2$,
for $\sqrt{ s_0}=4.5$ GeV (squares).
The solid line gives the parametrization of the QCDSR results  through Eq.
(\ref{exp}). }
\label{fig4}
\end{figure}

In Fig. \ref{fig4} we show, through the squares, the $Q^2$ dependence
of the form factor $g_{Z_{cs}\eta_c K^*}(Q^2)$, obtained using $M^2=5.0$ GeV$^2$.
As can be seen by Fig.~\ref{fig3}, other values of the
Borel mass, in  the range  $4.0 \leq M^2 \leq 10.0$ GeV$^2$,  give equivalent
results for the form factor.
The coupling constant is defined as the value of the form factor at the
meson pole \cite{Bracco:2011pg}. Therefore,  we need to extrapolate
the form factor to a region of $Q^2$ where the QCDSR is not valid. To do that
we parametrize the QCDSR results for
$g_{Z_{cs}\eta_c K^*}(Q^2)$ using a monopole form:

\beq
g_{Z_{cs}\eta_c K^*}(Q^2) = {g_1\over g_2 +Q^2}.
\label{mon}
\enq
The fit gives $g_1 = 78.35~ \GeV^{-2}$ and $ g_2=24.3~\GeV$.
In Fig. \ref{fig4} we also show, through the line, the fit of the QCDSR results,
using Eq.~(\ref{mon}). The coupling constant is obtained by
using Eq.~(\ref{mon}) and $Q^2=-m_{K^*}^2$:

\beq
g_{Z_{cs}\eta_c K^*}=g_{Z_{cs}\eta_c K^*}(-m^2_{K^*})=(3.4\pm0.3)~~\GeV.
\label{coupK}
\enq
The uncertainty in Eq.~(\ref{coupK}) comes from
variations in $s_0$, $\lambda_{Z_{cs}}$ and $m_c$ in the ranges given in Tables I
and II. Using this in Eq.~(\ref{dec}), and varying $m_{Z_{cs}}$ in the range
$m_{Z_{cs}}=(3.97\pm0.08)~\GeV$ we get

\beq
\Gamma(Z_{cs}^+\to \eta_c K^{*+})=(10.8\pm 6.2)~\MeV.
\label{width2}
\enq


Next we consider the decays $Z_{cs}^+\to D_s^+\bar{D}^{*0}$ and $Z_{cs}^+\to
D_s^{*+}\bar{D}^{0}$. Here we give only the expressions for $Z_{cs}^+\to
D_s^+\bar{D}^{*0}$. The expression for $Z_{cs}^+\to D_s^{*+}\bar{D}^{0}$, can be
easily obtained from the prior by exchanging the corresponding mesons masses
and condensates. As always the phenomenological side is obtained by considering
the contribution of the $Z_{cs},~D_s$ and $D^*$ mesons to the
correlation function in Eq.~(\ref{3po}):
\beq
\Pi_{\mu\al}^{(phen)} (p,\pli,q)={-i\lambda_{Z_{cs}} m_{D^*}f_{D^*}f_{D_s}
m^2_{D_s}~g_{Z_{cs}D^*D_s}(q^2)
\over (m_c+m_s)(p^2-m_{Z_{cs}}^2)({\pli}^2-m_{D^*}^2)(q^2-m_{D_s}^2)}
\left(-g_{\mu\lambda}+{\pli_\mu \pli_\lambda\over m_{D^*}^2}\right)
\left(-g_\alpha^\lambda+{p_\alpha p^\lambda\over m_{Z_{cs}}^2}\right)
+\cdots.
\lb{phen3}
\enq

Following \cite{Dias:2013xfa}, in the OPE side we consider only the CC diagrams
and we work with the $\pli_\alpha \pli_\mu$ structure. We get:
\beq
\Pi^{(OPE)}= {-i m_c\over48\sqrt{2}\pi^2}\left[{\mixs\over m_c^2-q^2}
\int_0^1 d\alpha{\alpha(2+\alpha)\over m_c^2-(1-\al){\pli}^2}
-{\mix\over m_c^2-{\pli}^2}
\int_0^1 d\alpha{\alpha(2+\alpha)\over m_c^2-(1-\al)q^2}\right].
\label{ope3}
\enq
\begin{figure}[h]
\centerline{\epsfig{figure=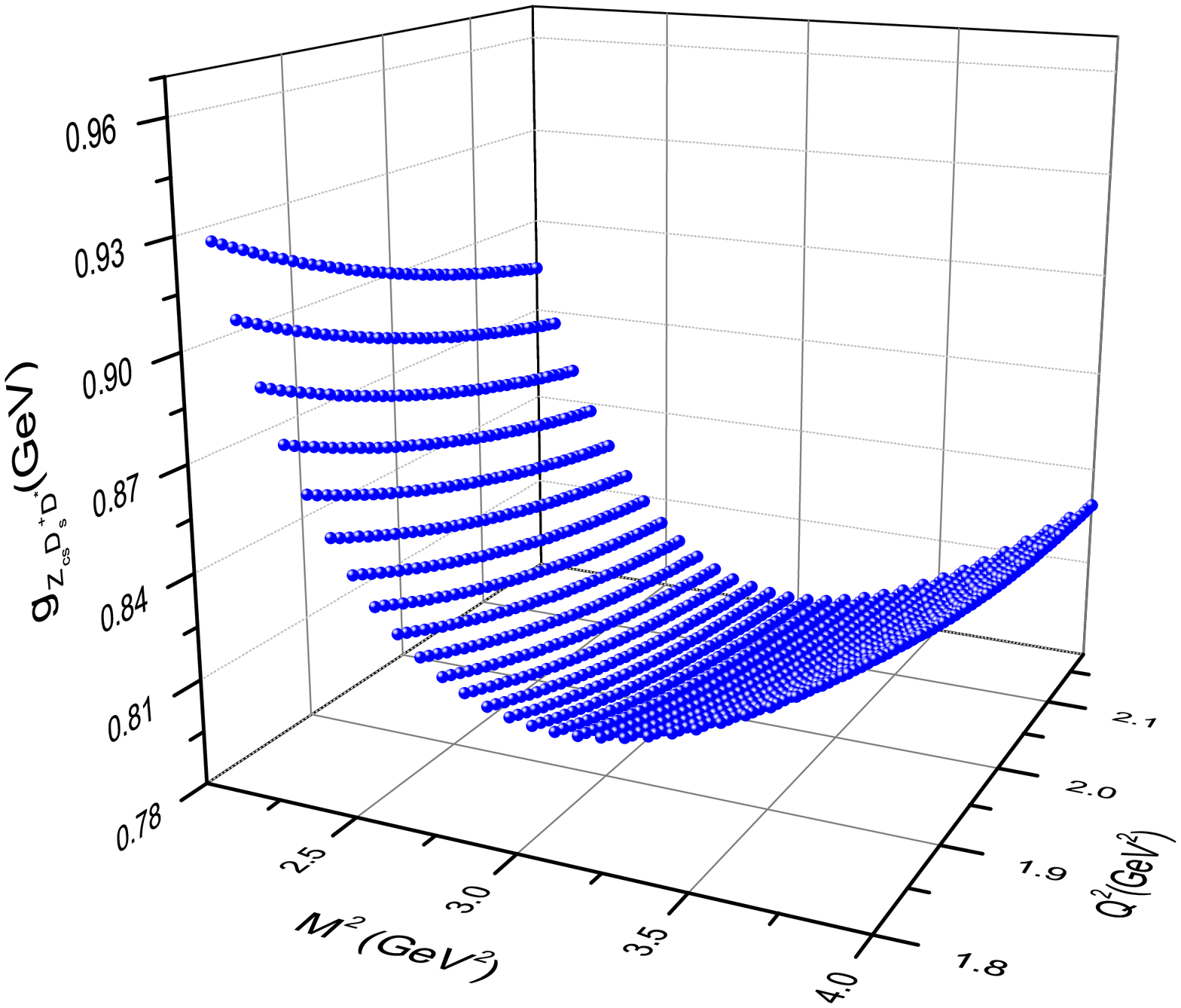,height=80mm}}
\caption{QCDSR results for the form factor  $g_{Z_{cs}D^*D_s}(Q^2)$ as
a function of $Q^2$ and $M^2$ for $\sqrt{ s_0}=4.5$ GeV.}
\label{fig5}
\end{figure}
Therefore, the sum rule in the $\pli_\mu\pli_\alpha$ structure is:
\beqa
&&{1\over Q^2+m_{D_s}^2}\left[E\left(e^{-m_{D^*}^2/M^2}-e^{-m_{Z_{cs}}^2/M^2}
\right)+F~e^{-s_0/M^2}\right]=
\nn\\
&&{m_c\over48\sqrt{2}\pi^2}\left[{\mixs\over m_c^2+Q^2}
\int_0^1 d\alpha{\alpha(2+\alpha)\over1-\al}~ e^{- m_c^2\over \al(1-\al)M^2}
-\mix e^{-m_c^2/M^2}
\int_0^1 d\alpha{\alpha(2+\alpha)\over m_c^2+(1-\al)Q^2}\right],
\label{sr3}
\enqa
where the parameter $E$ is defined in terms of the form factor $g_{Z_{cs}D_sD^*}
(Q^2)$:
\beq
E={g_{Z_{cs}D_sD^*}(Q^2)\lambda_{Z_{cs}} f_{D^*}f_{D_s}m_{D_s}^2
\over (m_c+m_s)m_{D^*}(m_{Z_{cs}}^2-m_{D^*}^2)}.
\label{e}
\enq
The form fator $g_{Z_{cs}D_sD^*}(Q^2)$ extracted from Eq.~(\ref{sr3}) is shown
in Fig. \ref{fig5}, as a function of both $M^2$ and $Q^2$. From this Fig. we
see that there is a good Borel stability in the region $2.75 \leq M^2 \leq 3.25$
GeV$^2$. Therefore, we fix $M^2=3.0~\GeV$ to extract the $Q^2$ dependence
of the form factor

\begin{figure}[h]
\centerline{\epsfig{figure=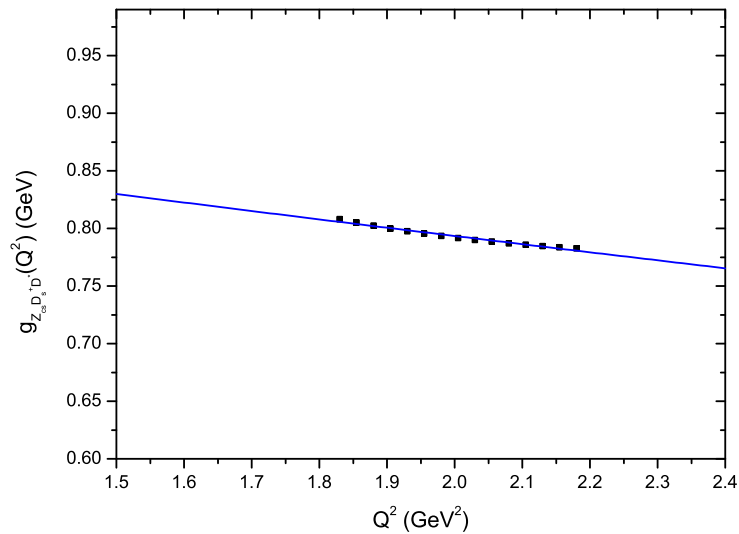,height=70mm}}
\caption{QCDSR results for $g_{Z_{cs}D^*D_s}(Q^2)$, as a function of $Q^2$,
for $\sqrt{ s_0}=4.5$ GeV (squares).
The solid line gives the parametrization of the QCDSR results  through Eq.
(\ref{exp}).}
\label{fig6}
\end{figure}


In Fig. \ref{fig6} we show, through the squares, the $Q^2$ dependence of the
form factor. Again, to extract the coupling constant we have to extrapolate
the QCDSR results to $Q^2=-m_{D_s}^2$. To do that we use an exponential form
\beq
g_{Z_{cs}D_sD^*}(Q^2) = g_1e^{-g_2 Q^2},
\label{exp}
\enq
to fit the QCDSR results. We have used an exponential form is this case  since
it was not possible to fit the QCDSR results with the monopole form in
Eq.~(\ref{mon}). However, as shown in \cite{Bracco:2011pg}, both forms are 
acceptable to describe hadronic form factors. We get
$g_1 = 0.94$ GeV and $ g_2=0.09~\mbox{GeV}^{-2}$.
The line in Fig. \ref{fig6} shows the fit of the QCDSR results
for $\sqrt{ s_0}=4.5$ GeV, using Eq.~(\ref{exp}).
We get for the coupling constant:

\beq
g_{Z_{cs}D_sD^*}=g_{Z_{cs}D_sD^*}(-m^2_{D_s})=(1.4 \pm 0.3)~~\mbox{GeV}.
\label{coupdd}
\enq
With this coupling and using the bigger value predicted for the $m_{Z_{cs}}$ mass
in \cite{Lee:2008uy} (since for values of the mass bellow the threshold the
decay is not possible) we get for the decay width in this channel:

\beq
\Gamma(Z_{cs}^+\to D_s^+\bar{D}^{*0})=(1.5\pm1.5)~\MeV.
\label{width3}
\enq

For the $Z_{cs}^+\to D_s^{*+}\bar{D}^{0}$, doing a similar analysis we arrive at:
\beq
g_{Z_{cs}D_s^*D}=g_{Z_{cs}D_s^*D}(-m^2_{D})=(1.4 \pm 0.4)~~\mbox{GeV},
\label{coupdd}
\enq
that leads to a similar result
\beq
\Gamma(Z_{cs}^+\to D_s^{*+}\bar{D}^{0})=(1.4\pm 1.4)~\MeV.
\label{width4}
\enq

\section{Conclusions}

In this work we have estimated, using the QCDSR approach, the decay widths
of the charmonium-like structure with hidden-charm and open-strange, that we
call $Z_{cs}^+$. This state was predicted in \cite{Chen:2013wca} under the ISChE
mechanism, and should be seen in the hidden-charm dikaon decay of  a
charmonium-like state $Y(4660)$. We have studied four decay channels and have
considered only color connected diagrams. This is justified by the fact that
we expect  the $Z_{cs}$ state to be a genuine tetraquark state, with a
non-trivial color configuration.
The obtained couplings, with the respective decay widths, are given in Table III.

\begin{center}
\small{{\bf Table III:} Coupling constants and decay widths in  different
channels.}
\\
\vskip3mm
\begin{tabular}{c|c|c}  \hline
 Vertex& coupling constant (GeV) & decay width (MeV)\\
\hline
 $Z_{cs}^+J/\psi K^+$ & $2.58\pm0.30$ & $11.2\pm3.5$ \\
 $Z_{cs}^+\eta_c K^{*+}$ & $3.4 \pm 0.3$ & $10.8\pm6.2 $ \\
 $Z_{cs}^+D_s^+\bar{D}^{*0}$ & $1.4\pm0.3$ & $1.5\pm1.5$ \\
 $Z_{cs}^+\bar{D}^0D_s^{*+}$ & $1.4 \pm0.4$ & $1.4\pm 1.4$ \\
\hline
\end{tabular}\end{center}

 Considering these
four decay channels we get a total width $\Gamma=( 24.9\pm 12.6)$ GeV for
$Z_{cs}$ which is  smaller than the total decay width of its
non-strange partner the $Z_c^+(3900)$:
$\Gamma=(46\pm 22)$ MeV from BESIII \cite{Ablikim:2013mio}, and
$\Gamma=(63\pm35)$ MeV from BELLE \cite{Liu:2013dau}.

 \subsection*{Acknowledgments}
 \noindent
This work has been supported by  CNPq and FAPESP-Brazil.
This work was also supported by the National Natural Science Foundation
of China under Grants 11222547, 11175073, 11035006,
 the Ministry of Education of China (FANEDD under Grant No. 200924,
SRFDP under Grant No. 20120211110002, NCET), the Fok Ying-Tong
Education Foundation (No. 131006).

\end{document}